\documentstyle[aps,prl,epsf]{revtex}
\begin{document}
\def\be{\begin{equation}}
\def\ee{\end{equation}}
\def\bea{\begin{eqnarray}}
\def\eea{\end{eqnarray}}
\def\oupb{UPB\ }
\def\pb{PB\ }
\def\oupbs{UPB's\ }
\def\pbs{PB's\ }
\newcommand{\ket}[1]{| #1 \rangle}
\newcommand{\bra}[1]{\langle #1 |}
\newcommand{\braket}[2]{\langle #1 | #2 \rangle}
\newcommand{\tr}{\rm tr}
\newcommand{\rank}{\rm rank}
\newcommand{\proj}[1]{| #1\rangle\!\langle #1 |}
\newcommand{\ba}{\begin{array}}
\newcommand{\ea}{\end{array}}
\newtheorem{theo}{Theorem}
\newtheorem{defi}{Definition}
\newtheorem{lem}{Lemma}
\newtheorem{exam}{Example}
\newtheorem{prop}{Property}
\newtheorem{coro}{Corollary}

\twocolumn[\hsize\textwidth\columnwidth\hsize\csname
@twocolumnfalse\endcsname

\author{Charles H. Bennett$^*$, David P. DiVincenzo$^*$, Tal Mor$^\dag$,
Peter W. Shor$^\ddag$, John A. Smolin$^*$, and Barbara M. Terhal$^\S$}

\title{Unextendible Product Bases and Bound Entanglement}

\address{\vspace*{1.2ex} \hspace*{0.5ex}{$^*$ IBM T.J. Watson Research
Center, Yorktown Heights, NY 10598, $^\dag$ DIRO, Universit\'e de
Montr\'eal, Canada and Dept. of Electrical Engineering, UCLA, Los
Angeles, CA 90095-1594, $^\ddag$AT\&T Labs--Research, Florham Park, NJ 07932,
$^\S$ ITF, UvA, Valckenierstraat 65, 1018 XE Amsterdam, and CWI,
Kruislaan 413, 1098 SJ Amsterdam, The Netherlands.\\ }}

\date{\today}

\maketitle
\begin{abstract}

An {\em unextendible product basis} (UPB) for a multipartite quantum
system is an incomplete orthogonal product basis whose complementary
subspace contains no product state. We give examples of UPBs, and show
that the uniform mixed state over the subspace complementary to any
UPB is a {\em bound entangled} state.
We exhibit a tripartite $2\!\times\!2\!\times\!2$
UPB whose complementary mixed state has tripartite entanglement but no
bipartite entanglement, i.e. all three corresponding $2\!\times\!4$
bipartite mixed states are unentangled. We show that members of a UPB
are not perfectly distinguishable by local POVMs and classical communication. 

\end{abstract}

]
%
 
Einstein, Podolsky and Rosen (EPR) first highlighted the necessarily nonlocal 
nature of quantum mechanics \cite{epr} which we now call
entanglement, that is, the paradox arising from the 
non-factorizability of the quantum states of two separated parties.
Bell showed that this entanglement implied a true nonlocality, or lack
of local realism in quantum mechanics \cite{bell}. 
But it is now clear that there are
other manifestations of quantum nonlocality that go beyond
entanglement \cite{qne}.  In this Letter we uncover various nonlocal
properties of some simple $m$-party sets of quantum states 
that involve only product (i.e., unentangled) states.

We present detailed examples of two-party and three-party sets of
orthogonal product states that are {\em unextendible}, meaning that no
further product states can be found orthogonal to all the existing
ones in a given Hilbert space.  We call such a set an unextendible
product basis (or UPB), and show that unextendibility gives rise to
two other recently discovered and not yet well-understood quantum
phenomena; 1) The mixed state on the subspace complementary to a UPB
is a {\em bound entangled} state
\cite{horodeckibound1,horodeckibound2}, i.e. a mixed state from which
no pure entanglement can be distilled.  We thus provide the first
systematic way of constructing bound entangled states, a task which
had been exceedingly difficult.  2) The states comprising a UPB are
{\em locally immeasurable}~\cite{qne}, i.e. an unknown member of the
set cannot be reliably distinguished from the others by local
measurements and classical communication.  Though sufficient,
unextendibility is not a necessary condition for either of these
phenomena, as there exist bound entangled states
\cite{horodeckibound1,horodeckibound2} not associated with any UPB,
and locally immeasurable sets of states which are not only extendible,
but capable of being extended all the way to a complete orthogonal
product basis on the entire Hilbert space \cite{qne}.

\begin{defi}
Consider a multipartite quantum system ${\cal H}=\bigotimes_{i=1}^m
{\cal H}_i$ with $m$ parties of respective dimension $d_i, i=1...m$. A
(incomplete orthogonal) {\em product basis (PB)\/} is a set $S$ of pure
orthogonal product states spanning a proper subspace ${\cal H}_S$ of
${\cal H}$. An {\em unextendible product basis (UPB)\/} is a {\rm PB}
whose complementary subspace ${\cal H}-{\cal H}_S$ contains no product
state. 
\end{defi}

To illustrate how a PB can fail to be extendible, consider the following
two sets of five states on $3 \!\times\! 3$ (two qutrits):

1) Let $\vec{v}_0,\vec{v}_1,\ldots,\vec{v}_4$ be five vectors in real 
three-dimensional space forming the apex of 
a regular pentagonal pyramid, the height $h$ of the pyramid being chosen 
such that nonadjacent vectors are orthogonal (cf.~Fig.~1). 
The vectors are  
\be
\vec{v}_j=N(\cos{{ 2 \pi j \over 5}},\sin{{2 \pi j}\over 5},h),\;\; 
j=0,\ldots,4,
\label{defP}
\ee
with $h={1 \over 2} \sqrt{1+\sqrt{5}}$ and $N=2/\sqrt{5+\sqrt{5}}$. 
Then the following five states in $3 \!\times\! 3$ Hilbert space 
form a UPB, henceforth denoted {\bf Pyramid}
\be
\ket{\psi_j}= \ket{\vec{v}_j} \otimes \ket{\vec{v}_{2j \bmod 5}}, \;\; j=0,\ldots,4.
\ee

To see that these five states form a UPB, note first that they are
mutually orthogonal: states whose indices differ by 2 mod 5 are
orthogonal for the first party (``Alice''); those whose indices differs by
1 mod 5 are orthogonal for the second party (``Bob''). For a new state
to be orthogonal to all the existing ones, it would have to be
orthogonal to at least three of Alice's states or at least three of
Bob's states. However this is impossible, since any three of the vectors
$\vec{v}_i$ span the full 3-dimensional space in which they live.  Therefore
the entire 4-dimensional subspace complementary to {\bf Pyramid} contains
no product state.  

2) The following five states on $3 \!\times\! 3$ form
a UPB henceforth denoted {\bf Tiles} 
\be 
\ba{lr} 
\ket{\psi_0}={1 \over
\sqrt{2}}\ket{0}(\ket{0}-\ket{1}),&\ \ \ket{\psi_2}= {1 \over
\sqrt{2}}\ket{2}(\ket{1}-\ket{2}),\\ \ket{\psi_1}={1 \over
\sqrt{2}}(\ket{0}-\ket{1})\ket{2},&\ \ \ket{\psi_3}= {1 \over
\sqrt{2}}(\ket{1}-\ket{2})\ket{0},\\ \ \ \ \
\lefteqn{\ket{\psi_4}=(1/3)(\ket{0}+\ket{1}+\ket{2})(\ket{0}+\ket{1}+\ket{2}).}  
\ea 
\ee
Note that the first four states are the interlocking tiles of
\cite{qne}, and the fifth state works as a ``stopper'' to force the
unextendibility.   

In both examples, any subset of three vectors on either side spans
the full three-dimensional space of that party, 
preventing any new vector from
being orthogonal to all the existing ones.
We formalize this observation by giving the necessary and sufficient 
condition for extendibility of a PB:

\begin{lem} 
Let $S=\{(\psi_j\!\equiv\!\bigotimes_{i=1}^m\varphi_{i,j}):j=1,\ldots,n\}$ 
be an incomplete orthogonal
product basis (PB) of an $m$-partite quantum system.  Let $P$ be a partition
of $S$ into $m$ disjoint subsets:
$S=S_1\cup  S_2\cup \ldots S_m$.  Let 
$r_i=\rank\{{\it \varphi_{i,j}:\psi_j\in S_i}\}$ be the local 
rank of subset $S_i$ as seen by the $i{\mbox th}$ party. 
Then $S$ is extendible iff there exists a partition $P$ such
that for all $i=1,\ldots,m$, the local rank $r_i$ of the 
$i{\mbox th}$ subset is less than the dimension $d_i$ of the 
$i{\mbox th}$ party's Hilbert space.  

\label{rule1}
\end{lem}

\noindent{\bf Proof:}$\;\;$ Imagine that the parties $i=1,\ldots,m$
allocate among themselves the job of being orthogonal to a new product
state we are trying to add. The new state will be orthogonal to all the
existing ones if a partition can be found such that the new state is
orthogonal to all the states in $S_1$ for party 1, all the states in
$S_2$ for party 2, and so on through $S_m$. Clearly this can be done
(e.g. by local Gramm-Schmidt orthogonalization for each party) if each
of the sets $S_i$ has local rank $r_i$ less than the dimensionality $d_i$ of
the its party's Hilbert space. Conversely, if for every partition, at
least one of the sets $S_i$ has full rank, equal to $d_i$, there is no
way to choose a new product state orthogonal to all the existing 
states; thus the original set is not extendible. $\Box$



The lemma provides a simple lower bound on the number of states $n$ in a 
UPB 
\be
n \geq \sum_i (d_i-1)+1,
\label{min}
\ee
since, for smaller $n$, one can partition S into sets of size $|{\rm S}_i|
\leq d_i-1$ and thus $r_i < d_i$ for all $m$ parties.


As noted earlier, UPBs provide a way to construct bound entangled (BE)
states, i.e.\ entangled mixed states from which no pure entanglement
can be distilled \cite{horodeckibound1,horodeckibound2}.  It was shown
in \cite{horodeckibound2} that if a bipartite density matrix $\rho$
remains positive semidefinite under the partial transposition
condition (PT) of Peres \cite{peresprl}, then $\rho$ cannot have distillable
entanglement. We then say that $\rho$ has positive partial
transposition (PPT).

\begin{theo}
The state that corresponds to the uniform mixture on the space
complementary to a UPB $\{\psi_i: i=1,\ldots,n\}$ in a Hilbert space of total
dimension $D$, 
\be
\bar{\rho}= {1 \over D-n}({\bf 1} -\sum_{j=1}^n \proj{\psi_j}),
\label{compl}
\ee
is a bound entangled state.
\label{rhobar}
\end{theo}

\noindent {\bf Proof:} By definition, the space complementary to a UPB
contains no product states.  Therefore $\bar\rho$ is entangled.
If the UPB is bipartite then $\bar\rho$ is PPT by construction: the identity 
is invariant under PT and the product states making up
the UPB are mapped onto another set of orthogonal product states.  
Therefore $PT(\bar\rho)$ is another density matrix, and thus positive semidefinite. 
For the case of many parties the PPT condition cannot be used directly,
so we use the above argument to show
that every bipartite partitioning of the parties is PPT.  Thus no
entanglement can be distilled across any bipartite cut.  If any
pure global entanglement could be distilled it could be used to
create entanglement across a bipartite cut.  Since $\bar\rho$ is entangled 
and is not distillable, it is bound entangled. $\Box$.


We compute that the state complementary to
the {\bf Tiles} UPB has a bound entanglement of formation\cite{bdsw} of
$0.213726$ ebits and the {\bf Pyramid} UPB has an entanglement of
formation of $0.232635$ ebits. These numbers are surprisingly large, 
considering that the maximal entanglement for any state in 
$3\!\times\!3$ is $\log_23\approx1.585$ ebits.

An example of a UPB involving three parties, 
$A$, $B$ and $C$, each holding a qubit, is the set 
\be
\{\ket{0,1,+},\ket{1,+,0},\ket{+,0,1},\ket{-,-,-}\},
\label{shift}
\ee
with $\pm=(\ket{0}\pm \ket{1})/\sqrt{2}$. 
One can see using Lemma \ref{rule1} that
there is no product state orthogonal to these four states, 
which we will henceforth call the {\bf Shifts} UPB.
This UPB can be simply generalized to a UPB over 
any number of parties, each with a one qubit Hilbert space (see \cite{ext}).

The complementary state to the {\bf Shifts} 
constructed by Eq. (\ref{compl}) has the curious property that not only 
is it two-way PPT, it is also two-way separable, i.e., the entanglement
across any split into two parties is zero.  This solves the main
problem left open in \cite{bras-mor} and surprisingly refutes a natural
 conjecture made there.  To show that the
entanglement between $A$ and $BC$ is zero, we write $a=\ket{1,+}$, 
$b=\ket{+,0}$, $c=\ket{0,1}$ and $d=\ket{-,-}$.  Note that these are just the
$B$ and $C$ parts of the four states in Eq. (\ref{shift}), and that
$\{a,b\}$ are orthogonal to $\{c,d\}$.
Consider the vectors $a^{\perp}$ and $b^{\perp}$ in the
$\mbox{span}(a,b)$ and the vectors $c^{\perp}$ and $d^{\perp}$ in the
$\mbox{span}(c,d)$. Now, we can complete the original set of vectors
to a full product basis between $A$ and $BC$ with the states
$\{\ket{0,a^{\perp}},\ket{1,b^{\perp}},\ket{+,c^{\perp}},\ket{-,d^{\perp}}\}$.
By the symmetry of the states, this is also true for the other splits. 

We now consider another nonlocal property of UPBs, their relation to
local immeasurability.  In this context, a useful notion, more general
than unextendibility, is {\em uncompletabilty}, an uncompletable
product basis being one that might be able to be extended by one or
more states, but cannot be completed to a full orthogonal product
basis on the entire Hilbert space.  Another concept we will need is
that of uncompletability even in a larger Hilbert space with local
extensions, where each party's space is extended from ${\cal H}_i$ to
${\cal H}_i \oplus {\cal H}_i'.$  We have the following simple fact
about UPBs:  

\begin{lem}
A UPB is not completable even in a locally extended Hilbert space.
\label{upbsareuncompletableinextendedspace}
\end{lem}

\noindent {\bf Proof:} If a set of states is completable in a locally
extended Hilbert space, then its complement space in the extended
Hilbert space is separable.  By local projections the uniform state on
the complementary space in the extended Hilbert space can be projected
onto the complementary space in the original Hilbert space.  This
state is also separable since local projections do not create
entanglement.  But we have a contradiction since the state complementary
to a UPB is entangled (Theorem \ref{rhobar}). $\Box$

We are now ready to show that uncompletability with local Hilbert
space extensions is a sufficient condition for a set of orthogonal
product states not to be perfectly distinguishable by any
sequence of local positive operator value measuements (POVMs) 
even with the help of classical communication among the observers.  
This form of local immeasurability was first studied in \cite{qne}.
To obtain a finite bound on the level of distinguishability of such
sets (as was done for the sets in \cite{qne}) requires additional
work and will be presented in \cite{ext}.

\begin{lem}
Given a set $S$ of orthogonal product states on ${\cal
H}=\bigotimes_{i=1}^m {\cal H}_i$ with $\mbox{dim } {\cal H}_i=d_i,
i=1...m$.  If the set $S$ is exactly measurable by local von Neumann
measurements and classical communication then it is completable in $\cal H$.
If $S$ is exactly measurable by local POVMs
and classical communication then the set can be completed in some 
extended space ${\cal H'}=\bigotimes_{i=1}^m ({\cal H}_i \oplus {\cal H}_i')$.
\label{immeasurability}
\end{lem}

\noindent {\bf Proof}: 
We show how a local von Neumann  measurement protocol 
leads directly to a way to complete the set $S$. At some stage of 
their protocol, the
parties (1) may have been able to eliminate members of the original
set of states $S$ and (2) they may have mapped, by performing their
von Neumann measurements, the remaining set of orthogonal states into
a new set of {\it orthogonal} states $S'$. Determining which
member they have in this new set uniquely determines with which state
of $S$ they started with.  At this stage, party $i_0$ performs an $l$-outcome 
von Neumann measurement which is given by a decomposition of the remaining
Hilbert space ${\cal K}={\cal K}_{\it else} \otimes {\cal K}_{i_0}$ with
${\cal K}_{\it else}=\bigotimes_{j \neq i_0}{\cal K}_j$, into a set of $l$
orthogonal subspaces, 
${\cal K}_{\it else} \otimes P_1 {\cal K}_{i_0}, ... 
{\cal K}_{\it else} \otimes P_l {\cal K}_{i_0}$

If a state in $S'$ lies in one of these subspaces, it will be unchanged
by the measurement. If a state $\ket{\alpha} \otimes \ket{\beta}$, 
where $\ket{\alpha} \in {\cal K}_{\it else}$, is not contained in one of 
the subspaces, it will be
projected onto one of the states $\{\ket{\alpha} \otimes P_1 \ket{\beta}, 
\ket{\alpha} \otimes P_2 \ket{\beta},\ldots,\ket{\alpha} 
\otimes P_l \ket{\beta}\}$. Let $S''$ be this new projected
set of states, containing both the unchanged states in $S'$ as well
as the possible projections of the states in $S'$.
If one of the subspaces does not contain a member of $S'$, it can be 
completed directly. For the
other subspaces, let us assume that each of them can be completed
individually with product states orthogonal to members of $S''$.  In
this way we have completed the projected $S'$ on the full Hilbert space
${\cal K}$, as these orthogonal-subspace completions are orthogonal
sets and they are a decomposition of ${\cal K}$.  However, we have now
completed the set $S''$ rather than the set $S'$.  Fortunately, one can
replace the projected states $\ket{\alpha} \otimes P_1 
\ket{\beta},\ldots,\ket{\alpha} \otimes P_l \ket{\beta}$ by the original 
state $\ket{\alpha} \otimes \ket{\beta}$ and $l-1$ orthogonal states by
making $l$ linear combinations of the projected states. 
They are orthogonal to
all other states as each $\ket{\alpha} \otimes P_i \ket{\beta}$ was 
orthogonal, and they can be made mutually orthogonal as they span an 
$l$-dimensional space on the $i_0$ side.  Thus at each round of 
measurement, a completion of the set of states $S'$ is achieved assuming 
a completion of the subspaces determined by the measurement.

The tree of nested subspaces will always lead to a subspace that
contains only a single state of the set, as the measurement protocol
was able to tell the states in $S$ apart exactly. But such a subspace
containing only one state can easily be completed and thus, by
induction, we have proved that the original set $S$ can be completed in 
${\cal H}$.

Finally, we note that a POVM is simply a von Neumann measurement in an
extended Hilbert space (this is Neumark's theorem (cf. \cite{peres}).
Thus any sequence of POVMs implementable locally with classical
communications is a sequence of local von Neumann measurements in extended
Hilbert spaces and the preceeding argument applies, leading to a completion
in $\cal H'$.  $\Box$

\begin{theo}
Members of a UPB are not perfectly distinguishable by local POVMs 
and classical communication.
\label{unmeas}
\end{theo}
\noindent{\bf Proof:} If the UPB were measurable by POVMs, it would be completable in some larger
Hilbert space by Lemma \ref{immeasurability}.  But this is in contradiction
with Lemma \ref{upbsareuncompletableinextendedspace}.  $\Box$

We now give an example of a PB that is measurable by local POVMs, but not by
local von Neumann measurements, which will also serve to illustrate the 
proofs just given.  The set is in a Hilbert space of dimension
$3 \!\times\! 4$. 
Consider the set $\vec{v}_j\otimes \vec{w}_j,\;\; j=0,\ldots,4$ with
$\vec{v}_j$ the states of the {\bf Pyramid} UPB as in Eq. (\ref{defP}) and
$\vec{w}_j$ defined as
\begin{eqnarray}
\nonumber\vec{w}_j&=N(\sqrt{\cos(\pi/5)} \cos(2j\pi/5),\sqrt{\cos(\pi/5)} \sin(2j\pi/5),\\
&\sqrt{\cos(2\pi/5)}\cos(4j\pi/5),\sqrt{\cos(2\pi/5)}\sin(4j\pi/5)),
\end{eqnarray}
with normalization $N=\sqrt{2/\sqrt{5}}$. Note that $\vec{w}_j^T
\vec{w}_{j+1}=0$ (addition mod $5$). One can show that this set,
albeit extendible on $3 \!\times\! 4$, is not {\it completable}: One
can at most add three vectors like $\vec{v}_0 \otimes
(\vec{w}_0,\vec{w}_1,\vec{w}_4)^{\perp}$, $\vec{v}_3 \otimes
(\vec{w}_2,\vec{w}_3,\vec{w}_4)^{\perp}$ and
$(\vec{v}_0,\vec{v}_3)^{\perp} \otimes
(\vec{w}_1,\vec{w}_2,\vec{w}_4)^{\perp}$.

The POVM measurement that is performed by Bob on the four-dimensional
side has five projector elements, each projecting onto a vector
$\vec{u}_j=N(-\sin(2 j \pi/5),\cos(2 j \pi/5),-\sin(4 j \pi/5),\cos(4j
\pi/5))$ with $j=0,\ldots,4$, and normalization $N=1/\sqrt{2}$.  Note
that $\vec{u}_0$ is orthogonal to vectors $\vec{w}_0,\vec{w}_2$ and
$\vec{w}_3$, or, in general, $\vec{u}_i$ is orthogonal to
$\vec{w}_i,\vec{w}_{i+2},\vec{w}_{i+3}$ (addition mod $5$). This means
that upon Bob's POVM measurement outcome, three vectors are excluded
from the set; then the remaining two vectors on Alice's side,
$\vec{v}_{i+1}$ and $\vec{v}_{i+4}$, are orthogonal and can thus be
distinguished.

The completion of this set is particularly simple. Bob's Hilbert space
is extended to a five-dimensional space. The POVM measurement can be
extended as a projection measurement in this five-dimensional space
with orthogonal projections onto the states
$\vec{x}_i=(\vec{u}_i,0)+\frac{1}{2}(0,0,0,0,1)$.  Then a completion
of the set in $3 \!\times\! 5$ are the following ten states:
\be
\ba{lr}
(\vec{v}_1,\vec{v}_4)^{\perp} \otimes \vec{x}_0, & \vec{v}_0 \otimes (\vec{w}_0^{\perp} \in \mbox{span}(\vec{x}_4,\vec{x}_1)), \\
(\vec{v}_0,\vec{v}_2)^{\perp} \otimes \vec{x}_1, & \vec{v}_1 \otimes (\vec{w}_1^{\perp} \in \mbox{span}(\vec{x}_0,\vec{x}_2)), \\
(\vec{v}_1,\vec{v}_3)^{\perp} \otimes \vec{x}_2, & \vec{v}_2 \otimes (\vec{w}_2^{\perp} \in \mbox{span}(\vec{x}_1,\vec{x}_3)), \\
(\vec{v}_2,\vec{v}_4)^{\perp} \otimes \vec{x}_3, & \vec{v}_3 \otimes (\vec{w}_3^{\perp} \in \mbox{span}(\vec{x}_2,\vec{x}_4)), \\
(\vec{v}_0,\vec{v}_3)^{\perp} \otimes \vec{x}_4, & \vec{v}_4 \otimes (\vec{w}_4^{\perp} \in \mbox{span}(\vec{x}_3,\vec{x}_0)). 
\ea
\ee

Although UPBs provide an easy way to construct a wide variety of bound entangled 
states, not all bound entangled states can be constructed by this
method.  In \cite{horodeckibound1} a $2 \!\times\! 4$ state with
bound entanglement was presented. However, our construction fails for
any $2 \!\times\! n$ as there is no UPB in $2 \!\times\! n$ for any
$n$. This follows from Theorem \ref{unmeas} and the fact that any set
of orthogonal product states on $2 \!\times\! n$ is locally
measurable.  The measurement is a three round protocol. Assume Alice
has the two-dimensional side. We write the set of states as $\{\ket{\alpha_1}
\otimes 
\ket{\beta_1},\ldots,\ket{\alpha_k} \otimes \ket{\beta_k}\}$. 
Alice can divide the states $\ket{\alpha_1},\ldots,\ket{\alpha_k}$ 
on her side in sets $P_i$ that have to be mutually
orthogonal on Bob's side, namely
$P_1=
\{\ket{\alpha_1},\ket{\alpha_1},\ldots,\ket{\alpha_1}^{\perp},\ldots\},\ldots,
P_i=\{\ket{\alpha_i},\ket{\alpha_i},\ldots,\ket{\alpha_i}^{\perp}\}$ etc. 
as $\ket{\alpha_i}$ and $\ket{\alpha_j}$ for $i \neq j$ are neither 
orthogonal nor identical. Note that $\ket{\alpha_i}$ or
$\ket{\alpha_i}^{\perp}$ can be repeated, but if they are, these
states should be orthogonal on Bob's side.  Now Bob can do a projection
measurement that singles out a subspace associated with $P_i$. After Bob sends
this information, the label $i$, to Alice, she does a measurement that
distinguishes $\ket{\alpha_i}$ from $\ket{\alpha_i}^{\perp}$. Then Bob 
can finish the protocol by measuring among the orthogonal repeaters.

Thus the following implications hold among properties of an incomplete,
orthogonal product basis:
Unextendible $\Rightarrow$
Complementary mixed state is BE $\Rightarrow$
Not completable even in a locally extended Hilbert space$\Rightarrow$
Not measurable exactly by local POVMs and classical communication $\Rightarrow$
Not measurable exactly by local von Neumann measurements and 
classical communication $\Rightarrow$
Preparation process $\ket{j}\rightarrow\ket{\psi_j}$ is
thermodynamically irreversible if carried out locally (cf.~\cite{qne}).

These constructions, relating entanglement to several other kinds of
nonlocality and irreversibility, may help illuminate some of the
following questions: How can BE states be used in quantum
protocols---for example, to perform otherwise non-local separable
superoperators~\cite{qne}? Can local hidden variable descriptions of
BE states be ruled out? 
What relation is there between the irreversibility implicit in the
definition of BE states and the thermodynamic
irreversibility~\cite{qne} of preparation of locally immeasurable sets
of states complementary to them?

We acknowledge the support of Elsag-Bailey and the Institute for Scientific 
Interchange during the 1998 Quantum Computation Workshop. CHB, DPD and JAS 
are grateful for support by the US Army Research Office under contract
\#DAAG55-98-C-0041 .   The work of TM was  supported in part by grant 
\#961360 from the Jet Propulsion Lab.
 
\begin{figure}
\epsfxsize=7cm
\epsfbox{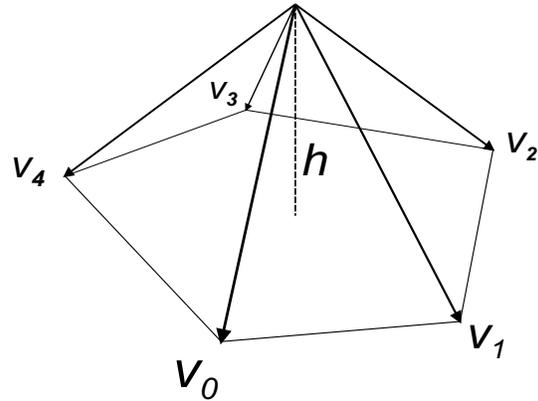}
\caption{{\bf Pyramid} vectors in real 3-space.  The height $h$ is chosen 
so that
$v_0 \perp v_{2,3}$ etc.}
\label{Pyramid}
\end{figure}


\begin{references}
\bibitem{epr}A. Einstein, B. Podolsky, N. Rosen, Phys.~Rev.~{\bf 47} (1935) 777.
\bibitem{bell}J.S. Bell, Physics (N.Y.) {\bf 1}, 195 (1964).
\bibitem{qne} C.H. Bennett, D.P. DiVincenzo, C.A. Fuchs, T. Mor,
E. Rains, P.W. Shor, J.A. Smolin, and W.K. Wootters, Phys. 
Rev. A {\bf 59}, 1070 (1999).
\bibitem{horodeckibound1} P. Horodecki, Phys. Lett. A {\bf232}, 333 (1997).  
\bibitem{horodeckibound2} M. Horodecki, P. Horodecki, and R. Horodecki,
Phys. Rev. Lett. {\bf 80}, 5239 (1998).
\bibitem{peresprl}A. Peres, Phys. Rev. Lett. {\bf 77}, 1413 (1996).
\bibitem{bdsw}C.H. Bennett, D.P. DiVincenzo, J.A. Smolin, and W.K. Wootters,
 Phys. Rev. A. {\bf 54}, 3824 (1996).
\bibitem{ext} C.H. Bennett, D.P. DiVincenzo, T. Mor, P.W. Shor, J.A. Smolin and
B.M. Terhal, in preparation.
\bibitem{bras-mor}G.~Brassard and T.~Mor, ``Multi-particle entanglement via
2-particle entanglement'', presented at NASA QCQC'98 conference.
\bibitem{peres} A. Peres, {\it Quantum Theory: Concepts and Methods}, Kluwer Academic
Publishers (1993).


\end{references}
\end{document}